\documentclass[a4paper,twocolumn,10pt,accepted=2022-04-16]{quantumarticle}
\pdfoutput=1
\usepackage[numbers]{natbib}

\usepackage[utf8]{inputenc}
\usepackage[english]{babel}
\usepackage[T1]{fontenc}
\usepackage{booktabs}

\usepackage{bbm}
\usepackage{amsbsy}
\usepackage{amsthm}
\usepackage{amssymb}
\usepackage{amsmath, mathtools}
\usepackage{dsfont}
\usepackage{xcolor}
\usepackage[colorlinks]{hyperref}
\makeatletter
\usepackage[figure,table]{hypcap}
\usepackage{enumerate}
\usepackage{braket}
\usepackage{soul}

\newcommand{\ketbra}[2]{\ensuremath{|{#1}\rangle\!\langle{#2}|}}
\newcommand{\nl}{\ensuremath{\hspace*{-0.5pt}}}
\newcommand{\subtiny}[3]{\ensuremath{_{\hspace{#1 pt}\protect\raisebox{#2 pt}{\tiny{$ #3$}}}}}
\newcommand{\suptiny}[3]{\ensuremath{^{\hspace{#1 pt}\protect\raisebox{#2 pt}{\tiny{$ #3$}}}}}

\DeclareMathOperator{\diag}{diag}

\newcommand{\tr}{\textnormal{Tr}}

\begin{document}
\title{Activation of genuine multipartite entanglement: Beyond the single-copy paradigm of entanglement characterisation}
\author{Hayata Yamasaki}
\email{hayata.yamasaki@oeaw.ac.at}
\affiliation{Institute for Quantum Optics and Quantum Information --- IQOQI Vienna, Austrian Academy of Sciences, Boltzmanngasse 3, 1090 Vienna, Austria}
\affiliation{Atominstitut,  Technische  Universit{\"a}t  Wien, Stadionallee 2, 1020  Vienna,  Austria}
\author{Simon Morelli}
\email{simon.morelli@oeaw.ac.at}
\affiliation{Institute for Quantum Optics and Quantum Information --- IQOQI Vienna, Austrian Academy of Sciences, Boltzmanngasse 3, 1090 Vienna, Austria}
\affiliation{Atominstitut,  Technische  Universit{\"a}t  Wien,  Stadionallee 2, 1020  Vienna,  Austria}
\author{Markus Miethlinger}
\affiliation{Institute for Quantum Optics and Quantum Information --- IQOQI Vienna, Austrian Academy of Sciences, Boltzmanngasse 3, 1090 Vienna, Austria}
\author{Jessica Bavaresco}
\affiliation{Institute for Quantum Optics and Quantum Information --- IQOQI Vienna, Austrian Academy of Sciences, Boltzmanngasse 3, 1090 Vienna, Austria}
\affiliation{Atominstitut,  Technische  Universit{\"a}t  Wien,  Stadionallee 2, 1020  Vienna,  Austria}
\author{Nicolai Friis}
\email{nicolai.friis@univie.ac.at}
\affiliation{Institute for Quantum Optics and Quantum Information --- IQOQI Vienna, Austrian Academy of Sciences, Boltzmanngasse 3, 1090 Vienna, Austria}
\affiliation{Atominstitut,  Technische  Universit{\"a}t  Wien,  Stadionallee 2, 1020  Vienna,  Austria}
\author{Marcus Huber}
\email{marcus.huber@univie.ac.at}
\affiliation{Atominstitut,  Technische  Universit{\"a}t  Wien,  Stadionallee 2, 1020  Vienna,  Austria}
\affiliation{Institute for Quantum Optics and Quantum Information --- IQOQI Vienna, Austrian Academy of Sciences, Boltzmanngasse 3, 1090 Vienna, Austria}

\begin{abstract}
    Entanglement shared among multiple parties presents complex challenges for the characterisation of different types of entanglement. One of the most fundamental insights is the fact that some mixed states can feature entanglement across every possible cut of a multipartite system yet can be produced via a mixture of states separable with respect to different partitions. To distinguish states that genuinely cannot be produced from mixing such partition-separable states, the term \emph{genuine multipartite entanglement} was coined. All these considerations originate in a paradigm where only a single copy of the state is distributed and locally acted upon. In contrast, advances in quantum technologies prompt the question of how this picture changes when multiple copies of the same state become locally accessible. Here we show that multiple copies unlock genuine multipartite entanglement from partially separable states, i.e., mixtures of the partition-separable states, even from undistillable ensembles, and even more than two copies can be required to observe this effect. With these findings, we characterise the notion of genuine multipartite entanglement in the paradigm of multiple copies and conjecture a strict hierarchy of activatable states and an asymptotic collapse of the hierarchy.
\end{abstract}

\maketitle

\section{Introduction}\label{sec:introduction}

Entanglement shared among multiple parties is acknowledged as one of the fundamental resources driving the second quantum revolution~\cite{DowlingMilburn2003}, for instance, as a basis of quantum network proposals~\cite{EppingKampermannMacchiavelloBruss2017, PivoluskaHuberMalik2018, RibeiroMurtaWehner2018, BaeumlAzuma2017}, as a key resource for improved quantum sensing~\cite{Toth2012} and quantum error correction~\cite{Scott2004} or as generic ingredient in quantum algorithms~\cite{BrussMacchiavello2011} and measurement-based quantum computation~\cite{RaussendorfBriegel2001, BriegelRaussendorf2001}. Yet, its detection and characterisation are complicated by several factors: among them, the computational hardness of deciding whether any given system even exhibits any entanglement at all~\cite{Gurvits2004} as well as the fact that the usual paradigm of local operations and classical communication (LOCC) lead to infinitely many types of entanglement~\cite{VerstraeteDehaeneDeMoorVerschelde2002, OsterlohSiewert2005, DeVicenteSpeeKraus2013, SchwaigerSauerweinCuquetDeVicenteKraus2015, DeVicenteSpeeSauerweinKraus2017, SpeeDeVicenteSauerweinKraus2017, SauerweinWallachGourKraus2018} already for single copies of multipartite states. Significant effort has thus been devoted to devising practical means of entanglement certification from limited experimental data~\cite{TothGuehne2005b, FriisVitaglianoMalikHuber2019}.

\begin{figure}[t]
\centering
\includegraphics[width=0.49\textwidth]{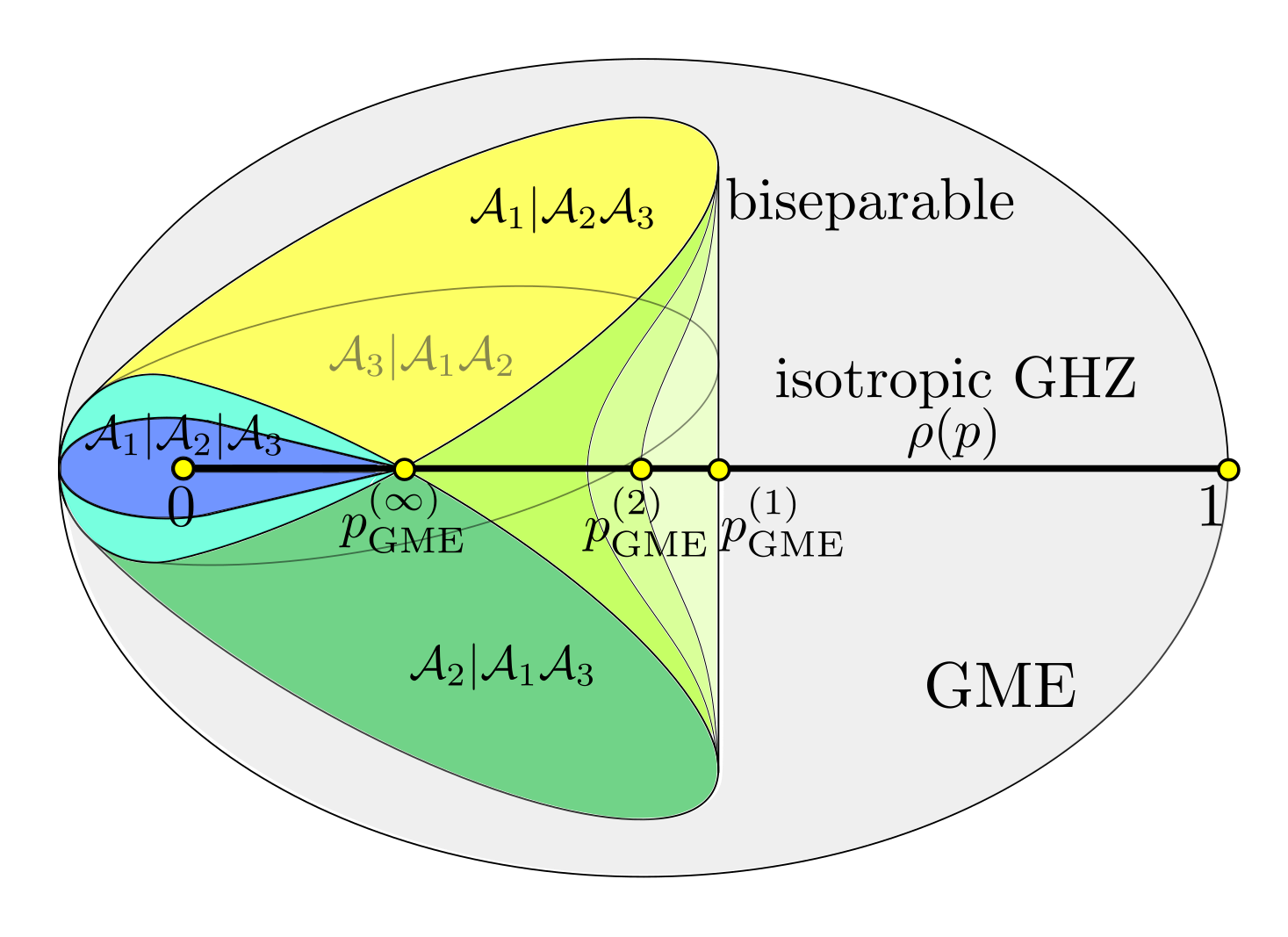}
\vspace*{-8mm}
\caption{\textbf{GME and (partial) separability for three qubits}. All three-qubit states separable with respect to~one of the three bipartitions, $\mathcal{A}_{1}|\mathcal{A}_{2}\mathcal{A}_{3}$ (yellow), $\mathcal{A}_{2}|\mathcal{A}_{1}\mathcal{A}_{3}$ (darker green), and $\mathcal{A}_{3}|\mathcal{A}_{1}\mathcal{A}_{2}$ (background), form convex sets, whose intersection (turquoise) contains (but is not limited to) all fully separable states $\mathcal{A}_{1}|\mathcal{A}_{2}|\mathcal{A}_{3}$ (dark blue). The convex hull of these partition-separable states contains all partially separable (the same as biseparable for tripartite systems) states. All states that are not biseparable are GME\@. States with $k$-copy activatable GME are contained in the set of biseparable but not partition-separable states and are conjectured to form the lighter green areas, with those states for which GME is activatable for higher values of $k$ farther away from the border between GME and biseparability.
The horizontal line represents the family of isotropic GHZ states $\rho(p)$, containing the maximally mixed state ($p=0$) and the GHZ state ($p=1$). The values $p\suptiny{0}{0}{(k)}_{\mathrm{GME}}$ indicate $k$-copy GME activation thresholds, which we discuss in the following.}
\label{fig:GME structure}
\end{figure}

One of the principal challenges for the characterisation of multipartite entanglement lies in distinguishing between \emph{partial separability} and its counterpart, \emph{genuine multipartite entanglement} (GME)\footnote{Note that the term was also coined for multipartite pure states with exclusively non-vanishing $n$-tangle in Ref.~\cite{OsterlohSiewert2005}.}.
Here, a multipartite state is called \emph{partially separable} if it can be decomposed as a mixture of \emph{partition-separable} states, i.e., of states separable with respect to some (potentially different) partitions of the parties into two or more groups, whereas any state that cannot be decomposed in this way has GME (see Fig.~\ref{fig:GME structure} and Table~\ref{tab:term}). One may further classify partially separable states as $k$-separable states according to the maximal number $k$ of tensor factors that all terms in the partially separable decomposition can be factorised into. If a state admits a decomposition where each term is composed of at least two tensor factors ($k=2$), the state is called \emph{biseparable}. Thus, every partially separable state is $k$-separable for some $k\geq2$, and hence (at least) biseparable.
This distinction arises naturally when considering the resources required to create a specific state:
any biseparable state can be produced via LOCC in setups where all parties share classical randomness and subsets of parties share entangled states.
One of the counter-intuitive features of partially separable states is the possibility for bipartite entanglement across every possible bipartition\footnote{An explicit example of a $k/2$-separable (and thus biseparable) $k$-qubit state (for even $k$) with the bipartite entanglement between all neighbouring qubits in a linear arrangement can be found in~\cite[footnote 30]{FriisMartyEtal2018}.}.
Consequently, the notion of bipartite entanglement across partitions is insufficient to capture the notion of partial separability, and conventional methods, such as positive maps~\cite{HorodeckiMPR1996, Peres1996}, cannot be straightforwardly applied to reveal GME (with new concepts for positive maps derived for that purpose in~\cite{HuberSengupta2014, ClivazHuberLamiMurta2017}), which results in additional challenges compared to the \textemdash~relatively \textemdash~simpler scenario of detecting bipartite or partition entanglement (e.g., as in~\cite{RodriguezBlancoBermudezMuellerShahandeh2021}).

\begin{figure}[t!]
\centering
\includegraphics[width=0.48\textwidth]{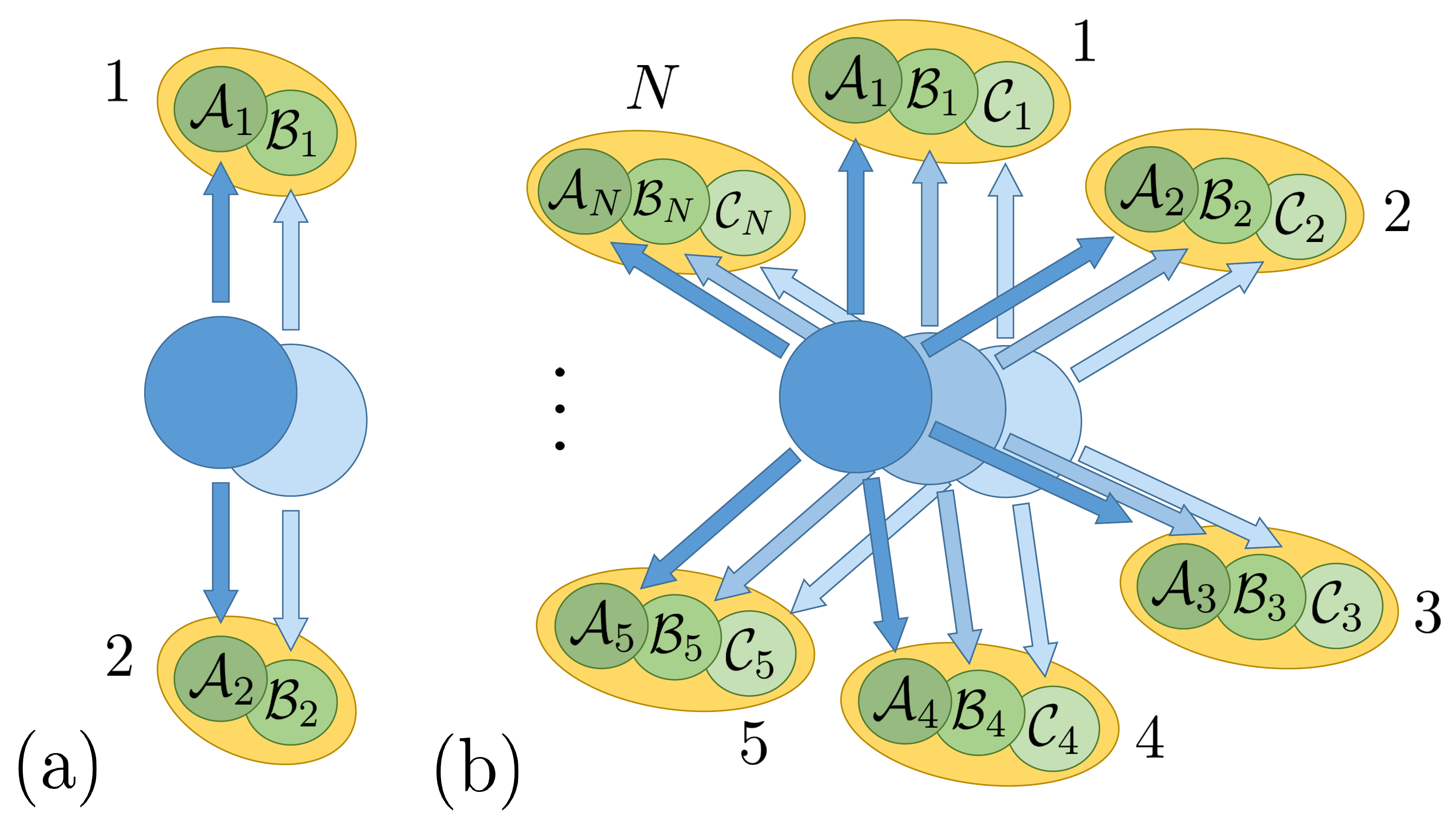}
\vspace*{-4mm}
\caption{\textbf{Activation of GME from biseparable states}. (a) Separable bipartite states remain separable, no matter how many copies are shared, e.g., if $\rho\subtiny{0}{0}{\mathcal{A}_{1}\mathcal{A}_{2}}$ and $\rho\subtiny{0}{0}{\mathcal{B}_{1}\mathcal{B}_{2}}$ are separable with respect to~the bipartitions $\mathcal{A}_{1}|\mathcal{A}_{2}$ and $\mathcal{B}_{1}|\mathcal{B}_{2}$, then so is  $\rho\subtiny{0}{0}{\mathcal{A}_{1}\mathcal{A}_{2}}\otimes\rho\subtiny{0}{0}{\mathcal{B}_{1}\mathcal{B}_{2}}$. (b) In contrast, the joint state of multiple copies of biseparable states, e.g., $\rho\subtiny{0}{0}{\mathcal{A}_{1},\mathcal{A}_{2},\ldots,\mathcal{A}_{N}}$, $\rho\subtiny{0}{0}{\mathcal{B}_{1},\mathcal{B}_{2},\ldots,\mathcal{B}_{N}}$, and $\rho\subtiny{0}{0}{\mathcal{C}_{1},\mathcal{C}_{2},\ldots,\mathcal{C}_{N}}$, can be GME with respect to~the partition $\mathcal{A}_{1}\mathcal{B}_{1}\mathcal{C}_{1}|\mathcal{A}_{2}\mathcal{B}_{2}\mathcal{C}_{2}|\ldots|\mathcal{A}_{\!N}\mathcal{B}_{\!N}\mathcal{C}_{\!N}$.
\label{fig:GME activation}
}
\end{figure}

\begin{table*}
\centering
\caption{\label{tab:term}Summary of terminology on GME in this paper.}
\begin{tabular}{@{}ll@{}}
\toprule
Term & Meaning\\
\midrule
$k$-separable &\parbox{12cm}{convex combination of pure states, each of which is a product of at least $k$ projectors}\\
biseparable & synonymous with $2$-separable\\
partially separable& $k$-separable for some $k>1$\\
partition-separable & \parbox{12cm}{separable for a specific partition of the multipartite Hilbert space, i.e., a convex combination of projectors, each of which is a product with respect to the same partition into subsystems}\\
multipartite entangled & entangled across all bipartitions\\
genuine multipartite entangled& non-biseparable\\
\bottomrule
\end{tabular}
\end{table*}

An assumption inherent in the definitions above is that all parties locally act only on a single copy of the distributed state. 
However, in many experiments where quantum states are distributed among (potentially distant) parties, multiple independent but identically prepared copies of states are (or at least, can be) shared. For instance, exceptionally high visibilities of photonic states can only be achieved if each detection event stems from almost identical quantum states~\cite{JoshieEtAl2020,WengerowskyJoshiSteinlechnerHuebelUrsin2018}. Adding noise to the channel then produces the situation we focus on in this article: multiple copies of noisy quantum states produced in a laboratory~\cite{Ecker-Huber2019,HuEtAl2020}.
Even limited access to quantum memories or signal delays then allows one to act on multiple copies of the distributed states, which is a recurring theme also in research on quantum networks~\cite{YamasakiPirkerMuraoDuerKraus2018,NavascuesWolfeRossetPozasKerstjens2020, KraftDesignolleRitzBrunnerGuehneHuber2021}.
Characterising properties of GME in multi-copy scenarios is thus not only of fundamental theoretical interest but also crucial for practical applications that require GME to be distributed, such as conference key agreement~\cite{MurtaGrasselliKampermannBruss2020}.

However, we demonstrate here that, unlike the distinction between separable and entangled states, the distinction between biseparability and GME is not maintained in the transition from one to many copies; i.e., partial separability is not a tensor-stable concept.
As we show, for $N$ parties $1,\ldots,N$, there exist multipartite quantum states $\rho\subtiny{0}{0}{\mathcal{A}_{1},\mathcal{A}_{2},\ldots,\mathcal{A}_{N}}$ that are biseparable, but which can be \emph{activated} in the sense that sharing two copies results in a GME state, i.e., such that the joint state $\rho\subtiny{0}{0}{\mathcal{A}_{1},\mathcal{A}_{2},\ldots,\mathcal{A}_{N}}\otimes \rho\subtiny{0}{0}{\mathcal{B}_{1},\mathcal{B}_{2},\ldots,\mathcal{B}_{N}}$ of two identical copies (labelled $\mathcal{A}$ and $\mathcal{B}$, respectively) is not biseparable with respect to the partition $\mathcal{A}_{1}\mathcal{B}_{1}|\mathcal{A}_{2}\mathcal{B}_{2}|\ldots|\mathcal{A}_{N}\mathcal{B}_{N}$. (See Fig.~\ref{fig:GME activation}.)
That such activation of GME is in principle possible had previously only been noted in~\cite{HuberPlesch2011}, where it was observed that two copies of a particular four-qubit state that is itself almost fully separable can become GME\@. 
Here, we systematically investigate this phenomenon of \emph{multi-copy GME activation}. As the first main result, we show that the property of biseparability is not tensor stable in general by identifying a family of $N$-qubit isotropic Greenberger-Horne-Zeilinger (GHZ) states with two-copy activatable GME for all $N$.
We further demonstrate the existence of biseparable states within this family for which two copies are not enough to activate GME, but three copies are.
Moreover, we show that the bound for partition-separability coincides with the asymptotic (in terms of the number of copies) GME-activation bound for isotropic GHZ states.

Multi-copy GME activation is particularly remarkable \textemdash~and may appear surprising at first \textemdash~because it is in stark contrast to bipartite entanglement:
Two copies of states separable with respect to a fixed partition always remain partition-separable and can never become GME\@.
However, from the perspective of entanglement distillation \textemdash~the concentration of entanglement from many weakly entangled (copies of) states to few strongly entangled ones \textemdash~such an activation seems more natural.
After all, if one party shares bipartite maximally entangled states with each other party, these could be used to establish any GME state among all $N$ parties via standard teleportation, thus distributing GME by sharing only two-party entangled states.
Nevertheless, such a procedure would require at least $N-1$ copies of these bipartite entangled states (in addition to a local copy of the GME state to be distributed), and already the example from~\cite{HuberPlesch2011} suggests that one does not have to go through first distilling bipartite entangled pairs, followed by teleportation, but two copies can naturally feature GME already.
While we have seen that the phenomenon of GME activation is more than just distillation, one may still be tempted to think that distillable entanglement is required for GME activation.
It is known that there exist bound entangled states \textemdash~entangled states that do not admit distillation of entanglement no matter how many copies are provided.
In particular, all entangled states with positive partial transpose (PPT) across a given cut are undistillable since any number of copies is also PPT\@.
One might thus suspect that GME activation should not be possible for biseparable states that are PPT across every cut and hence have no distillable entanglement (even if multiple parties are allowed to collaborate).
As another main result, we show that this is not the case by constructing a biseparable state that is PPT with respect to~every cut, yet two copies of the state are indeed GME\@. 
Together, our results thus support the following conjectures:
\begin{enumerate}[(i)]
\item{\label{conjecture i}
There exists a hierarchy of states with $k$-copy activatable GME, i.e., for all $k\geq2$ there exists a biseparable but not partition-separable state $\rho$ such that $\rho^{\otimes k-1}$ is biseparable, but $\rho^{\otimes k}$ is GME\@.
}
\item{\label{conjecture ii}
GME may be activated for any biseparable but not partition-separable state (light green areas in Fig.~\ref{fig:GME structure}) of any number of parties as $k\rightarrow\infty$.}
\end{enumerate}

In the following, we first provide the formal definitions for biseparability and GME in Sec.~\ref{sec:sep and gme} before turning to the family of $N$-qubit isotropic GHZ states in Sec.~\ref{sec:GME of isotropic GHZ states}. For all biseparable states in this family, we provide upper bounds on the minimal number of copies required to activate GME in Sec.~\ref{sec:Multi-copy GME criterion}. In Sec.~\ref{sec:Hierarchy of k-copy activatable states}, we then consider the case of three qubits ($N=3$), for which we can show that the bound on three-copy GME activation is tight in the sense that we identify all states in the family for which one requires at least three copies to activate GME, while two copies remain biseparable, and can also show that GME can indeed be activated for any biseparable but not partition-separable state in this family. Moreover, in Sec.~\ref{sec:GME activation of PPT entangled states}, we construct an explicit example for two-copy GME activation from biseparable states with no distillable bipartite entanglement. Finally, we discuss the implications of our results and open questions in Sec.~\ref{sec:Conclusion and Outlook}.

\section{Definitions of biseparability \& GME}\label{sec:sep and gme}
We summarise the formal definitions of biseparability and GME in this paper.
(See also Table~\ref{tab:term} for the summary of the definitions here.)
Formally, a pure quantum state of an $N$-partite system with Hilbert space $\mathcal{H}\suptiny{0}{0}{(N)}=\bigotimes_{i=1}^{N}\mathcal{H}_{i}$ is separable with respect to~a $k$-partition $\{\mathcal{A}_{1},\mathcal{A}_{2},\ldots,\mathcal{A}_{k}\}$, with $\mathcal{A}_{i}\subset
\{1,2,3,\ldots,N\}$ and $\bigcup_{i=1}^{k} \mathcal{A}_{i}= \{1,2,3,\ldots,N\}$ such that $\bigotimes_{i=1}^{k}\mathcal{H}_{\mathcal{A}_{i}}=\mathcal{H}\suptiny{0}{0}{(N)}$, if it can be written as
\begin{align}
    \ket{\Phi\suptiny{0}{0}{(k)}}  &=\,\bigotimes\limits_{i=1}^{k}\,\ket{\phi_{\mathcal{A}_{i}}},\quad\ket{\phi_{\mathcal{A}_{i}}}\in\mathcal{H}_{\mathcal{A}_{i}}
    \,.\label{pure}
\end{align}
When generalising to density matrices, it is common not to specify all possible partitions, but to use the notion of \emph{$k$-separability} instead: 
A density operator is called \emph{$k$-separable} if it can be decomposed as a convex sum of pure states that are all separable with respect to~\emph{some} $k$-partition, i.e., if it is of the form (see, e.g., the review~\cite{FriisVitaglianoMalikHuber2019})
\begin{align}
    \rho\suptiny{0}{0}{(k)} &=\,
    \sum\limits_{i} p_{i} 
    \ket{\Phi_i\suptiny{0}{0}{(k)}}\!\!\bra{\Phi_i\suptiny{0}{0}{(k)}}
    \,.\label{ksep}
\end{align}
Note that the lack of tensor stability of partial separability shown in the following also implies that the related concept of $k$-producibility~\cite{GuehneToth2009,Szalay2019} is not tensor stable. Crucially, each $\ket{\Phi_i\suptiny{0}{0}{(k)}}$ may be $k$-separable with respect to~a different $k$-partition. Consequently, $k$-separability does not imply separability of $\rho\suptiny{0}{0}{(k)}$ with respect to~a specific partition, except when $\rho\suptiny{0}{0}{(k)}$ is a pure state or when $k=N$. In the latter case the state is called \emph{fully separable}.
To make this distinction more explicit, we refer to all (at least) biseparable states that are actually separable with respect to~some bipartition as \emph{partition-separable}.
At the other end of this separability spectrum one encounters \emph{biseparable states} ($k=2$), while all states that are not at least biseparable (formally, $k=1$) are called \emph{genuinely $N$-partite entangled}. We will here use the term GME for the case $k=1$.
The operational reason for this definition of GME is easily explained: any biseparable state of the form of Eq.~(\ref{ksep}) can be created by $N$ parties purely by sharing partition-separable states of the form of Eq.~(\ref{pure}) and some classical randomness. 
In addition, this conveniently results in a convex notion of biseparability (as illustrated for the example in Fig.~\ref{fig:GME structure}) amenable to entanglement witness techniques, which inherently rely on convexity.

\section{GME of isotropic GHZ states}\label{sec:GME of isotropic GHZ states}
To overcome the difficulty in analysing GME, the crucial technique here is to use states in $X$-form, i.e., those with nonzero entries of density operators only on the main diagonal and main anti-diagonal with respect to~the computational basis.
Let us now consider a family of mixed $N$-qubit states, \emph{isotropic GHZ states}, given by
\begin{align}
    \rho(p) &=\,p\,\ket{\mathrm{GHZ}_{N}\!}\!\!\bra{\mathrm{GHZ}_{N}\!}\,+\,(1-p)\,\tfrac{1}{2^{N}}\mathds{1}_{2^{N}}\,,
    \label{eq:GHZ with white noise}
\end{align}
obtained as convex combination of the $N$-qubit maximally mixed state $\tfrac{1}{2^{N}}\mathds{1}_{2^{N}}$ and a pure 
$N$-qubit GHZ state
\begin{align}
    \ket{\mathrm{GHZ}_{N}\!}  &=\,\tfrac{1}{\sqrt{2}}\bigl(\ket{0}^{\otimes N}+\ket{1}^{\otimes N}\bigr).
\end{align}
with real mixing parameter $p\in[-1/(2^{N}-1),1]$.
Since states in this family are in $X$-form with respect to~the $N$-qubit computational basis, we can straightforwardly calculate the \emph{genuine multipartite} (GM) \emph{concurrence}, an entanglement measure for a multipartite state defined in terms of a polynomial of elements of its density matrix~\cite{HashemiRafsanjaniHuberBroadbentEberly2012,MaChenChenSpenglerGabrielHuber2011}. For any $N$-qubit density operator $\rho_{X}$ in $X$-form, i.e., 
\begin{align}
    \rho_{X}=\begin{pmatrix} \tilde{a} & \tilde{z}\,\tilde{d} \\ \tilde{d}\,\tilde{z}^{\dagger} & \tilde{d}\,\tilde{b}\,\tilde{d} \end{pmatrix},
\end{align}
where $\tilde{a}=\diag\{a_{1},\ldots,a_{n}\}$,  $\tilde{b}=\diag\{b_{1},\ldots,b_{n}\}$, and $\tilde{z}=\diag\{z_{1},\ldots,z_{n}\}$ are diagonal $n\times n$ matrices with $n=2^{N-1}$, $a_{i},b_{i}\in\mathbb{R}$ and $z_{i}\in\mathbb{C}$ for all $i=1,2,\ldots,n$, and $\tilde{d}=\operatorname{antidiag}\{1,1,\ldots,1\}$ is antidiagonal,
the GM concurrence is given by
\begin{align}
    C_{\mathrm{GM}}(\rho_{X}) &=\,2\max\bigl\{0,\max_{i}\{|z_{i}|-\sum\limits_{j\neq i}^{n}\sqrt{a_{j}b_{j}}\}\bigr\},
    \label{eq:GM concurrence}
\end{align}
and provides a necessary and sufficient condition for GME whenever $C_{\mathrm{GM}}>0$. 
In the case of the state $\rho(p)$ from Eq.~(\ref{eq:GHZ with white noise}), we have $a_{i}=b_{i}=\tfrac{1-p}{2^{N}}+\delta_{i1}\tfrac{p}{2}$ and $z_{i}=\delta_{i1}\tfrac{p}{2}$, such that
\begin{align}
    C_{\mathrm{GM}}\bigl[\rho(p)\bigr] &=\,\max\{0,|p|-(1-p)(1-2^{1-N})\}.
\end{align}
Thus, $\rho(p)$ is GME if and only if
\begin{align}
    p   &>\,p\suptiny{0}{0}{(1)}_{\mathrm{GME}}(N)\,\coloneqq\,\frac{2^{N-1}-1}{2^{N}-1}\,,
\end{align}
i.e., if and only if $p$ surpasses the single-copy threshold $p\suptiny{0}{0}{(1)}_{\mathrm{GME}}$.
Conversely, we can be certain that $\rho(p)$ is not GME for $p\leq (2^{N-1}-1)/(2^{N}-1)$, and hence at least biseparable.

\section{Multi-copy GME criterion}\label{sec:Multi-copy GME criterion}
Our first goal is then to check if two copies of $\rho(p)$ are GME\@.
Since the GM concurrence is an entanglement monotone, $C_\mathrm{GM}\bigl[\rho(p)^{\otimes k}\bigr]$ is monotonically non-decreasing as $k$ increases~\cite{MaChenChenSpenglerGabrielHuber2011};
that is, if we have $C_\mathrm{GM}\bigl[\rho(p)\bigr]=0$ for $\rho(p)$ in $X$-form, it holds that $C_\mathrm{GM}\bigl[\rho(p)^{\otimes 2}\bigr]\geq 0$ in general.
However, 
using $C_\mathrm{GM}\bigl[\rho(p)^{\otimes 2}\bigr]> 0$
as a necessary and sufficient criterion for GME 
is not an option in this case, 
since ${\rho(p)}^{\otimes 2}$ may not be of $X$-form even if a single copy is, and we therefore generally cannot directly calculate $C_\mathrm{GM}\bigl[\rho(p)^{\otimes 2}\bigr]$. 
The crucial idea here is to make use of the fact that stochastic LOCC (SLOCC) can never create GME from a biseparable state.

To construct a sufficient GME criterion,
we therefore use a map $\mathcal{E}_{\circ}$ implementable via SLOCC~\cite{LamiHuber2016}, which, for any two density operators $\rho$ and $\sigma$ acting on $\mathcal{H}$,
maps the state $\rho\otimes\sigma$  acting on $\mathcal{H}^{\otimes 2}$ to
\begin{align}
    \mathcal{E}_{\circ}[\rho\otimes\sigma] &=\,\frac{\rho\circ\sigma}{\tr(\rho\circ\sigma)}\quad\text{on }\mathcal{H},
\end{align}
where the right-hand side is a density operator acting on $\mathcal{H}$, and ``$\circ$'' denotes the Hadamard product (or Schur product), i.e., the component-wise multiplication of the two matrices.
What is useful for us here is that the Hadamard product of two $X$-form matrices results in an $X$-form matrix. Consequently, we can directly calculate the GM concurrence for the state resulting from applying the `\emph{Hadamard-product map}' $\mathcal{E}_{\circ}$ to two copies of an originally biseparable state.
If the GM concurrence of $\mathcal{E}_{\circ}[\rho(p)^{\otimes 2}]$ is nonzero, we can conclude that two copies of $\rho(p)$ are GME, even if a single copy is not.
To decide whether $\mathcal{E}_{\circ}[\rho(p)^{\otimes 2}]$ is GME or not, i.e., whether the GM concurrence is nonzero or not, we can ignore the normalization and just consider $\rho(p)\circ\rho(p)=\rho(p)^{\circ 2}$. Moreover, in the maximization over the index $i$ in Eq.~(\ref{eq:GM concurrence}), the maximum is obtained for $i=1$. We can thus conclude that $\rho(p)^{\otimes 2}$ is GME if
\begin{align}
    |z_{1}^{2}|-\sum\limits_{j\neq 1}^{n}\sqrt{a_{j}^{2}b_{j}^{2}}\,=\,\tfrac{p^{2}}{4}-(2^{N-1}-1)\bigl(\tfrac{1-p}{2^{N}}\bigr)^{2}\,>\,0,
    \label{eq:nonzero GM concurrence 2 copies}
\end{align}
which translates to the condition $p/(1-p)>\sqrt{2^{N-1}-1}/2^{N-1}$, and in turn can be reformulated to the condition
\vspace*{-2mm}
\begin{align}
    p   &>\,p\suptiny{0}{0}{(2)}_{\mathrm{GME}}(N)\,\coloneqq\,\frac{\sqrt{2^{N-1}-1}}{2^{N-1}+\sqrt{2^{N-1}-1}}.
    \label{eq:GME treshhold 2 copies}
\end{align}
As we see, we have $p\suptiny{0}{0}{(1)}_{\mathrm{GME}}>p\suptiny{0}{0}{(2)}_{\mathrm{GME}}$ for all $N\geq3$, confirming that \emph{there exist biseparable states} with values $p<p\suptiny{0}{0}{(1)}_{\mathrm{GME}}$ \emph{for which two copies are GME}, i.e., such that $p>p\suptiny{0}{0}{(2)}_{\mathrm{GME}}$. 

Moreover, we can now concatenate multiple uses of the SLOCC map $\mathcal{E}_{\circ}$. For instance, we can identify the threshold value $p\suptiny{0}{0}{(3)}_{\mathrm{GME}}$ of $p$ at which the state $\mathcal{E}_{\circ}\bigl[\rho(p)\otimes\mathcal{E}_{\circ}[\rho(p)^{\otimes 2}]\bigr]$ resulting from $2$ applications of $\mathcal{E}_{\circ}$ to a total of $3$ copies of $\rho(p)$ is GME, or, more generally, the corresponding threshold value $p\suptiny{0}{0}{(k)}_{\mathrm{GME}}$ for which $k$ copies result in a GME state after applying the map $\mathcal{E}_{\circ}$ a total of $k-1$ times. From Eq.~(\ref{eq:nonzero GM concurrence 2 copies}) it is easy to see that these threshold values are obtained as
\begin{align}
    p\suptiny{0}{0}{(k)}_{\mathrm{GME}}(N)\,\coloneqq\,\frac{\sqrt[k]{2^{N-1}-1}}{2^{N-1}+\sqrt[k]{2^{N-1}-1}}.
\end{align}

\section{Hierarchy of $k$-copy activatable states}\label{sec:Hierarchy of k-copy activatable states}
The threshold values $p\suptiny{0}{0}{(k)}_{\mathrm{GME}}$ provide upper bounds on the minimal number of copies required to activate GME\@: a value $p$ satisfying $p\suptiny{0}{0}{(k)}_{\mathrm{GME}}<p<p\suptiny{0}{0}{(k-1)}_{\mathrm{GME}}$ for $k\geq2$ implies that $k$ copies are enough to activate GME\@.
But since the map $\mathcal{E}_{\circ}$ (does not create and) may reduce GME, it does not imply that $k$ copies are actually needed; up to this point, there is a possibility that two copies are all it takes.

However, at least for the case of three qubits ($N=3$) and three copies ($k=3$), we find that this is not the case. That is, for all isotropic three-qubit GHZ states with $p\le p\suptiny{0}{0}{(2)}_{\mathrm{GME}}(N=3)=\sqrt{3}/(4+\sqrt{3})$, we find that two copies are still biseparable, and thus at least three copies are required to activate GME\@. The explicit biseparable decomposition of two copies of the states in this range is presented in Appendix~\ref{sec:appendix two copy bisep}. Although it does not constitute conclusive proof, this result nevertheless supports our first conjecture, repeated here for convenience:

\noindent
\emph{Conjecture~(\ref{conjecture i}):\ There exists a hierarchy of states with $k$-copy activatable GME, i.e., for all $k\geq2$ there exists a biseparable but not partition-separable state $\rho$ such that $\rho^{\otimes k-1}$ is biseparable, but $\rho^{\otimes k}$ is GME\@.}

The conjectured existence of a hierarchy of biseparable states with $k$-copy activatable GME means that states become less and less `valuable' as the number of copies $k$ required to obtain GME increases. At the same time, it is also clear that all partition-separable states cannot be used to activate GME because separability with respect to~any fixed partition is tensor stable. But it is not clear where exactly the boundary between activatable and non-activatable states really lies (see Fig.~\ref{fig:GME structure}).

To shed light on this question, let us again examine the isotropic GHZ states from Eq.~(\ref{eq:GHZ with white noise}) with regards to partition-separability with respect to~the bipartition separating the first qubit $\mathcal{A}_{1}$ from the remaining $N-1$ qubits (collected in $\tilde{\mathcal{A}}_{2}$), i.e., $\mathcal{A}_{1}|\tilde{\mathcal{A}}_{2}$. Using this partition, we can write
\begin{align}
    \rho(p) &=p\ket{\Phi^{+}}\!\!\bra{\Phi^{+}}_{\mathcal{A}_{1}\tilde{\mathcal{A}}_{2}}
    \!+\!\tfrac{1-p}{2^{N}}\,\mathds{1}_{\!\mathcal{A}_{1}}\hspace*{-3pt}\otimes\!
    \mathds{1}_{\!\tilde{\mathcal{A}}_{2}}
    \!+\!\tfrac{1-p}{2^{N}}\,\mathds{1}_{\!\mathcal{A}_{1}}\hspace*{-3pt}\otimes\!\mathds{1}_{\!\tilde{\mathcal{A}}_{2}^{\perp}}\nonumber\\
    &=\,
    \tfrac{1+p}{2}\,\tilde{\rho}_{\mathcal{A}_{1}\tilde{\mathcal{A}}_{2}}\,
    +\tfrac{1-p}{2}\,\tfrac{1}{2^{N-1}}\,
    \mathds{1}_{\mathcal{A}_{1}}\hspace*{-3pt}\otimes\!
    \mathds{1}_{\tilde{\mathcal{A}}_{2}^{\perp}},
    \label{eq:GHZ with white noise rewritten}
\end{align}
where $\ket{\Phi^{+}}_{\mathcal{A}_{1}\tilde{\mathcal{A}}_{2}}=\tfrac{1}{\sqrt{2}}\bigl(\ket{0}_{\mathcal{A}_{1}}\ket{\tilde{0}}_{\tilde{\mathcal{A}}_{2}}+\ket{1}_{\mathcal{A}_{1}}\ket{\tilde{1}}_{\tilde{\mathcal{A}}_{2}}\bigr)$ with $\ket{\tilde{i}}_{\tilde{\mathcal{A}}_{2}}=\bigotimes_{j=2}^{N}\ket{i}_{\mathcal{A}_{j}}$ for $i=0,1$, $\mathds{1}_{\!\tilde{\mathcal{A}}_{2}}=\sum_{i=0,1}\ket{\tilde{i}}\!\!\bra{\tilde{i}}$ and $\mathds{1}_{\!\tilde{\mathcal{A}}_{2}^{\perp}}=\mathds{1}_{2^{N-1}}-\mathds{1}_{\!\tilde{\mathcal{A}}_{2}}$. From this decomposition, it becomes clear that the state can be written as a convex sum of a two-qubit state $\tilde{\rho}_{\mathcal{A}_{1}\tilde{\mathcal{A}}_{2}}$ (where the second qubit lives on the two-dimensional subspace of $\tilde{\mathcal{A}}_{2}$, spanned by the states $\ket{\tilde{i}}$ for $i=0,1$) and diagonal terms proportional to $\mathds{1}_{\mathcal{A}_{1}}\otimes\mathds{1}_{\!\tilde{\mathcal{A}}_{2}^{\perp}}$ with support in a subspace $\tilde{\mathcal{A}}_{2}^{\perp}$ orthogonal to $\tilde{\rho}_{\mathcal{A}_{1}\tilde{\mathcal{A}}_{2}}$. The latter diagonal terms trivially have a separable decomposition with respect to~the bipartition $\mathcal{A}_{1}|\tilde{\mathcal{A}}_{2}$. For the two-qubit state $\tilde{\rho}_{\mathcal{A}_{1}\tilde{\mathcal{A}}_{2}}$, the PPT criterion offers a necessary and sufficient separability criterion, and one easily finds that the partial transpose of $\tilde{\rho}_{\mathcal{A}_{1}\tilde{\mathcal{A}}_{2}}$ is non-negative if $p\leq p_{\mathrm{crit}}\coloneqq 1/(1+2^{N-1})$ (see Appendix~\ref{appendix:PPT criterion for isotropic GHZ states}). Further taking into account its qubit exchange symmetry, we thus find that $\rho(p)$ is partition-separable with respect to~any bipartition for $p\leq p_{\mathrm{crit}}$. At the same time, we find that $\lim_{k\rightarrow\infty}p\suptiny{0}{0}{(k)}_{\mathrm{GME}}(N)=p_{\mathrm{crit}}$, which implies that any isotropic GHZ state with $p>p_{\mathrm{crit}}$ features $k$-copy activatable GME, at least asymptotically as $k\rightarrow\infty$, and is thus also not partition-separable. This leads us to our second conjecture, also repeated here for convenience:

\noindent
\emph{Conjecture~(\ref{conjecture ii}):\ 
GME may be activated for any biseparable but not partition-separable state of any number of parties as $k\rightarrow\infty$.}

Conjecture~(\ref{conjecture ii}) holds for isotropic GHZ states. But does it hold in general?

\section{GME activation from PPT entangled states}\label{sec:GME activation of PPT entangled states}
A situation where one might imagine Conjecture~(\ref{conjecture ii}) to fail is the situation of biseparable (but not partition-separable) states with PPT entanglement across every bipartition, as discussed in Sec.~\ref{sec:introduction}.
For isotropic GHZ states, however, the PPT criterion across every cut coincides exactly with the threshold $p_{\mathrm{crit}}$ for biseparability (and GME activation), as one can confirm by calculating the eigenvalues of the partial transpose of $\rho(p)$ (see Appendix~\ref{appendix:PPT criterion for isotropic GHZ states}).
We thus turn to a different family of states, for which this is not the case.

Specifically, as we show in detail in Appendix~\ref{appendix:PPT entangled GME activation}, we construct a family of biseparable three-party states 
\begin{align}
    \rho_{\mathcal{A}_{1}\mathcal{A}_{2}\mathcal{A}_{3}
    } &=\,
    \sum\limits_{\substack{i,j,k=1\\ i\neq j\neq k\neq i}}^{3}p_{i}\ \rho_{\mathcal{A}_{i}}\otimes\rho_{\mathcal{A}_{j}\mathcal{A}_{k}}\suptiny{0}{0}{\mathrm{PPT}}
\end{align}
where the $\rho_{\mathcal{A}_{j}\mathcal{A}_{k}}\suptiny{0}{0}{\mathrm{PPT}}$ are (different) two-qutrit states with PPT entanglement across the respective cuts $\mathcal{A}_{j}|\mathcal{A}_{k}$ for $j\neq k\in\{1,2,3\}$ and $\sum_{i}p_{i}=1$. Via LOCC, three copies (labelled $\mathcal{A}$, $\mathcal{B}$, and $\mathcal{C}$, respectively) of this state $\rho_{\mathcal{A}_{1}\mathcal{A}_{2}\mathcal{A}_{3}}$ can be converted to what we call \emph{PPT-triangle states} of the form
\begin{equation}
    \rho_{\mathcal{A}_{2}\mathcal{A}_{3}}\suptiny{0}{0}{\mathrm{PPT}}\otimes \rho_{\mathcal{B}_{1}\mathcal{B}_{3}}\suptiny{0}{0}{\mathrm{PPT}}\otimes \rho_{\mathcal{C}_{1}\mathcal{C}_{2}}\suptiny{0}{0}{\mathrm{PPT}}.
\end{equation}
Using a GME witness based on the lifted Choi map (cf.~\cite{HuberSengupta2014, ClivazHuberLamiMurta2017}), we show that there exists a parameter range where these PPT-triangle states are GME.
Therefore, it is proved that GME activation is possible even from biseparable states only with PPT entanglement across every bipartition.

\section{GME activation and shared randomness}

Provided that our conjectures are true, incoherent mixing (access to shared randomness) can lead to situations where the number of copies needed for GME activation is reduced. In the extreme case, and this is true even based only on the results already proven here (and thus independently of whether or not the conjectures turn out to be true or not), the probabilistic combination of partition-separable states (without activatable GME) can results in a state \textemdash\ a biseparable isotropic GHZ state \textemdash\ which has activatable GME. Although this may at first glance appear to be at odds with the usual understanding of bipartite entanglement, which cannot arise from forming convex combinations of separable states, we believe this can be understood rather intuitively if we view incoherent mixing as a special case of a more general scenario in which one may have any amount of information on the states that are shared between different observers. As an example, consider the following situation:

Three parties, labelled, $1$, $2$ and $3$, share two identical (as in, the system and its subsystems have the same Hilbert space dimensions and are represented by the same physical degrees of freedom) tripartite quantum systems, labelled $\mathcal{A}$ and $\mathcal{B}$, in the states $\rho_{\mathcal{A}_{1}|\mathcal{A}_{2}\mathcal{A}_{3}}$ and $\rho_{\mathcal{B}_{1}\mathcal{B}_{2}|\mathcal{B}_{3}}$, respectively, where we assume that $\rho_{\mathcal{A}_{1}|\mathcal{A}_{2}\mathcal{A}_{3}}$ is separable with respect to the bipartition $\mathcal{A}_{1}|\mathcal{A}_{2}\mathcal{A}_{3}$ and $\rho_{\mathcal{B}_{1}\mathcal{B}_{2}|\mathcal{B}_{3}}$ is separable with respect to the bipartition $\mathcal{B}_{1}\mathcal{B}_{2}|\mathcal{B}_{3}$. Clearly, both of these systems and states individually are biseparable, but if the parties have full information about which system is which, e.g., the first system is $A$ and the second system is $B$, then the joint state $\rho_{\mathcal{A}_{1}|\mathcal{A}_{2}\mathcal{A}_{3}}\otimes\rho_{\mathcal{B}_{1}\mathcal{B}_{2}|\mathcal{B}_{3}}$ can be GME with respect to the partition $\mathcal{A}_{1}\mathcal{B}_{1}|\mathcal{A}_{2}\mathcal{B}_{2}|\mathcal{A}_{3}\mathcal{B}_{3}$. In this sense, two biseparable systems can yield one GME system. Now, let us suppose that the parties do not have full information which system is in which state. For simplicity, let us assume that either system may be in either state with the same probability $\tfrac{1}{2}$. Then the state of either of the systems is described by the convex mixture $\rho_{\mathrm{mix}}=\tfrac{1}{2}\rho_{A_{1}|A_{2}A_{3}}+\tfrac{1}{2}\rho_{B_{1}B_{2}|B_{3}}$, where we have kept the labels $A$ and $B$, but they now refer to the same subsystems, i.e., $A_{i}=B_{i}$ for all~$i$. The state $\rho_{\mathrm{mix}}$ may not be partition separable anymore, but is certainly still biseparable. In particular, it may have activatable GME, even though neither $\rho_{A_{1}|A_{2}A_{3}}$ nor $\rho_{B_{1}B_{2}|B_{3}}$ do. For the sake of the argument let us assume that the latter is indeed the case and that GME is activated for $2$ copies in this case, such that $\rho_{\mathrm{mix}}^{\otimes 2}$ is GME. That means, if one has access to both systems, $A$ and $B$, even without knowing which system is in which state, one would end up with GME. However, the additional randomness with respect to the case where one knows exactly which state which system is in results in an increased entropy of $\rho_{\mathrm{mix}}^{\otimes 2}$ with respect to $\rho_{\mathcal{A}_{1}|\mathcal{A}_{2}\mathcal{A}_{3}}\otimes\rho_{\mathcal{B}_{1}\mathcal{B}_{2}|\mathcal{B}_{3}}$, and thus represents a disadvantage with respect to the latter case.

In general, it is therefore not problematic that the conjectures, if true, would imply that incoherent mixtures of $k$-activatable states may result in $k'$-activatable states with $k'<k$. Instead, this can be considered as a sign that scenarios with multiple copies of multipartite quantum states give rise to features that are not captured by convex structures on the level of the single-copy state space.

\section{Conclusion and outlook}\label{sec:Conclusion and Outlook}
Our results show that a modern theory of entanglement in multipartite systems, which includes the potential to locally process multiple copies of distributed quantum states, exhibits a rich structure that goes beyond the convex structure of partially separable states on single copies. While we conjecture that asymptotically, an even simpler description might be possible, i.e., separability in multipartite systems collapses to a simple bipartite concept of separability, we show that two copies are certainly not sufficient for reaching this simple limit, thus leaving the practical certification with finite copies a problem to be studied.

Indeed, our results show that GME is a resource with a complex relationship to bipartite entanglement in the context of local operations and shared randomness (cf.~\cite{SchmidRossetBuscemi2020}). An array of important open questions arises from our results, which can thus be considered to establish an entirely new direction of research: first and foremost, this includes the quest for conclusive evidence for or against our conjectures. Besides determining whether these conjectures are ultimately correct or not, it will be of high interest to determine which properties (of the biseparable decompositions) of given states permit or prevent GME activation with a certain number of copies. Another open question is the minimal local dimension necessary for GME activation from biseparable states with PPT entanglement across every cut. Furthermore, from a practical point of view, it will be desirable to develop a theory of $k$-copy multipartite entanglement witnesses that are non-linear expressions of density matrices and allow for a more fine-grained characterisation of multipartite entanglement in networks with local memories.
Finally, although separable states and shared classical randomness are free under LOCC, i.e., under a conventional choice of free operations in the resource theory of bipartite entanglement, our results suggest that convex combinations of different partition-separable states with shared classical randomness can be used as a resource for GME activation in multi-copy scenarios; that is, it may not be straightforward to study GME activation within the usual resource-theoretical framework under LOCC\@. 
In view of this situation, it would be interesting for future research to establish a new framework for understanding such a complicated aspect of multipartite entanglement as GME activation by, e.g., considering non-convex quantum resource theories where classical randomness can be used as a resource~\cite{Kuroiwa2020generalquantum,PhysRevA.104.L020401}.

\begin{acknowledgments}
We acknowledge support from the Austrian Science Fund (FWF) through the START project Y879-N27, the project P 31339-N27, and the Zukunftskolleg ZK03. H.Y.\ was supported by JSPS Overseas Research Fellowships and JST, PRESTO Grant Number JPMJPR201A, Japan.
\end{acknowledgments}

\bibliographystyle{apsrev4-1fixed_with_article_titles_full_names_new}
\bibliography{Master_Bib_File}

\onecolumngrid

\appendix
\section*{Appendix}

The appendices are organised as follows.
In Appendix~\ref{sec:appendix two copy bisep}, we analyse which values of the parameter allow for a biseparable decomposition of two copies of the three-qubit isotropic Greenberger-Horne-Zeilinger (GHZ) states.
In Appendix~\ref{appendix:PPT criterion for isotropic GHZ states}, we study the  positive-partial-transpose (PPT) criterion for isotropic GHZ states.
In Appendix~\ref{appendix:PPT entangled GME activation}, we show that multi-copy activation of genuine multipartite entanglement (GME) is possible from PPT bound entanglement.

\section{Biseparable decomposition of two-copy three-qubit isotropic GHZ states}\label{sec:appendix two copy bisep}

In this appendix we analyse which values of the parameter $p$ allow for a biseparable decomposition of two copies of the three-qubit isotropic GHZ states. To be more precise, we look for a biseparable decomposition with respect to~the partition $\mathcal{A}_{1}\mathcal{B}_{1}|\mathcal{A}_{2}\mathcal{B}_{2}|\mathcal{A}_{3}\mathcal{B}_{3}$ of the state $\rho_3(p)^{\otimes2}$, where
\begin{align}
    \rho_3(p) =\,p\,\ket{\mathrm{GHZ}_{3}}\!\!\bra{\mathrm{GHZ}_{3}}\,+\,(1-p)\,\tfrac{1}{2^{3}}\mathds{1}_{2^{3}}
\end{align}
is the three-qubit isotropic GHZ state defined in the main text.

To construct a biseparable decomposition, we first construct separable states for two or four qubits. We then map these states to different six-qubit states in such a way that all resulting six-qubit states are separable with respect to one of the bipartitions
\begin{equation}
\label{eq:decomposition}
    \mathcal{A}_{1}\mathcal{B}_{1}|\mathcal{A}_{2}\mathcal{B}_{2}\mathcal{A}_{3}\mathcal{B}_{3}, \mathcal{A}_{1}\mathcal{B}_{1}\mathcal{A}_{2}\mathcal{B}_{2}|\mathcal{A}_{3}\mathcal{B}_{3}, \mathcal{A}_{2}\mathcal{B}_{2}|\mathcal{A}_{1}\mathcal{B}_{1}\mathcal{A}_{3}\mathcal{B}_{3}.
\end{equation}
For convenience of notation, we henceforth reorder the subsystems to 
$\mathcal{A}_{1}\mathcal{B}_{1}\mathcal{A}_{2}\mathcal{B}_{2}\mathcal{A}_{3}\mathcal{B}_{3}$.
We then group together these different states to define biseparable states for the whole six-qubit system. This allows us to rewrite the state $\rho_3(p)^{\otimes2}$ as a convex sum of these biseparable states and a diagonal matrix. Finally, we find conditions for which this diagonal matrix has only non-negative entries, i.e., is positive semi-definite and thus itself a state. 

Let us begin by defining the separable two-qubit state
\begin{align}
   \gamma= \frac{1}{4}(\ketbra{++}{++}+\ketbra{--}{--}+\ketbra{rl}{rl}+\ketbra{lr}{lr}),
\end{align}
where $\Ket{+}=(\Ket{0}+\Ket{1})/\sqrt{2}$, $\Ket{-}=(\Ket{0}-\ket{1})/\sqrt{2}$, $\Ket{r}=(\Ket{0}-i\ket{1})/\sqrt{2}$ and $\Ket{l}=(\Ket{0}+i\ket{1})/\sqrt{2}$.
We partition the six-qubit space $\mathcal{A}_{1}\mathcal{B}_{1}\mathcal{A}_{2}\mathcal{B}_{2}\mathcal{A}_{3}\mathcal{B}_{3}$
into two subsystems $\mathcal{C}$ and $\mathcal{D}$ in such a way that the bipartition $\mathcal{C}|\mathcal{D}$ coincides with one of the three bipartitions in~\eqref{eq:decomposition}.
We then define a map $E$ from a two-qubit state space to the six-qubit space $\mathcal{A}_{1}\mathcal{A}_{2}\mathcal{A}_{3}\mathcal{B}_{1}\mathcal{B}_{2}\mathcal{B}_{3}$ as the unique linear map such that $\ket{00}\rightarrow\ket{ii'}$, $\ket{01}\rightarrow\ket{ij'}$, $\ket{10}\rightarrow\ket{ji'}$ and $\ket{11}\rightarrow\ket{jj'}$, where $\ket{i}$ and $\ket{j}$ are orthogonal states of subsystem $\mathcal{C}$ and $\ket{i'}$ and $\ket{j'}$ are orthogonal states of subsystem $\mathcal{D}$.
Applying this map to the two-qubit separable state $\gamma$ above, we have a six-qubit state $E(\gamma)$ that is separable across the cut $\mathcal{C|D}$ by construction.
In the following we will consider only such embeddings $E$ that map $\ket{00}$, $\ket{01}$, $\ket{10}$ and $\ket{11}$ onto four of the standard-basis states of the six-qubit space.
For example consider the partition $\mathcal{A}_{1}\mathcal{B}_{1}\mathcal{A}_{2}\mathcal{B}_{2}|\mathcal{A}_{3}\mathcal{B}_{3}$ and the embedding $E$ that maps $\ket{00}\rightarrow\ket{000000}$, $\ket{01}\rightarrow\ket{000001}$, $\ket{10}\rightarrow\ket{010100}$ and $\ket{11}\rightarrow\ket{010101}$. The embedded state then reads
\begin{align}
\label{eq:example}
    E(\gamma)=\frac{1}{4}(&\ketbra{000000}{000000}+\ketbra{000000}{010101}+\ketbra{010101}{000000}+\ketbra{010101}{010101}\notag\\&+\ketbra{000001}{000001}+\ketbra{010100}{010100}).
\end{align}

For every index $m$ running from $1$ to $64$, we let $\ket{m}=\ket{i_1i_2i_3i_4i_5i_6}$ denote a standard-basis state of $\mathcal{A}_{1}\mathcal{B}_{1}\mathcal{A}_{2}\mathcal{B}_{2}\mathcal{A}_{3}\mathcal{B}_{3}$ such that
\begin{equation}
\label{eq:m}
    m=32i_1+16i_2+8i_3+4i_4+2i_5+i_6+1,
\end{equation}
that is, $m$ is the decimal representation of the number represented by the bit string $i_1i_2i_3i_4i_5i_6$. 
Let $E_{m_1,m_2,m_3,m_4}$ be the linear map from a two-qubit space to the previously considered six-qubit space such that
\begin{subequations}
\begin{align}
 \ket{00}\mapsto \ket{m_1},\\
 \ket{01}\mapsto \ket{m_2},\\
 \ket{10}\mapsto \ket{m_3},\\
 \ket{11}\mapsto \ket{m_4}.
\end{align}
\end{subequations}
We then define
\begin{align}
    \gamma(m_1,m_2,m_3,m_4)=E_{m_1,m_2,m_3,m_4}(\gamma).
\end{align}
For example, the state~\eqref{eq:example} is denoted by $\gamma(1,2,21,22)=E_{1,2,21,22}(\gamma)$.
Note that not all combinations $m_1,m_2,m_3,m_4$ define a two-qubit subspace across the bipartitions in~\eqref{eq:decomposition}, and among all subspaces, we are only interested in those pertaining to different parties.
With this notation, we introduce the following states
\begin{align}
\begin{split}
    \Gamma_1 =\frac{1}{24}[ 
    &\gamma(2,10,36,44)+\gamma(2,12,34,44)
    +\gamma(33,37,50,54)+\gamma(3,7,20,24)+\gamma(3,8,19,24)\\
    +&\gamma(5,7,45,47)+\gamma(5,15,37,47)
    +\gamma(9,10,29,30)+\gamma(9,14,25,30)+\gamma(18,20,58,60)\\[1mm]
    +&\gamma(18,28,50,60)+\gamma(41,45,58,62)+\gamma(41,46,57,62)
    +\gamma(21,29,55,63)+\gamma(21,31,53,63)\\[1mm]
    +&\gamma(35,36,55,56)+\gamma(35,40,51,56)+\gamma(6,8,46,48)
    +\gamma(6,14,40,48)+\gamma(11,12,31,31)\\[1mm]
    +&\gamma(11,15,28,32)+\gamma(17,19,57,59)
    +\gamma(17,25,51,59)+\gamma(33,34,53,54)],
\end{split}
\end{align}

\begin{align}
\begin{split}
    \Gamma_2= \frac{1}{12}[
    &\gamma(1,2,21,22)+\gamma(1,5,18,22)
    +\gamma(1,6,17,22)+\gamma(1,3,41,43)+\gamma(1,9,35,43) \\
    +&\gamma(1,11,33,43)+\gamma(22,24,62,64)
    +\gamma(22,30,56,64)+\gamma(22,32,54,64)+\gamma(43,44,63,64)\\[1mm]
    +&\gamma(43,47,60,64)+\gamma(43,48,59,64)].
\end{split}
\end{align}

With the same notation as before we define the four-qubit separable state
\begin{align}
\begin{split}
   \sigma =\frac{1}{16}(
   &\ketbra{++++}{++++}\,+\,\ketbra{+-+-}{+-+-}
   +\ketbra{-+-+}{-+-+}\,+\,\ketbra{----}{----}\\
   +&\ketbra{+r+l}{+r+l}\,+\,\ketbra{+l+r}{+l+r}
   +\ketbra{-r-l}{-r-l}\,+\,\ketbra{-l-r}{-l-r}\\[1mm]
   +&\ketbra{r+l+}{r+l+}\,+\,\ketbra{r-l-}{r-l-}
   +\ketbra{l+r+}{l+r+}\,+\,\ketbra{l-r-}{l-r-}\\[1mm]
   +&\ketbra{rrll}{rrll}\,+\,\ketbra{rllr}{rllr}
   +\ketbra{lrrl}{lrrl}\,+\,\ketbra{llrr}{llrr}),
\end{split}
\end{align}
shared between three parties. It can be split in three different ways: $\sigma_{\mathcal{A}_{1}\mathcal{B}_{1}\mathcal{A}_{2}\mathcal{A}_{3}}$, $\sigma_{\mathcal{A}_{1}\mathcal{A}_{2}\mathcal{B}_{2}\mathcal{A}_{3}}$ and $\sigma_{\mathcal{A}_{1}\mathcal{A}_{2}\mathcal{A}_{3}\mathcal{B}_{3}}$. 
Next, we define the biseparable six-qubit state 
\begin{equation}
    \Sigma = \frac{1}{3}(U_1\sigma_{\mathcal{A}_{1}\mathcal{B}_{1}\mathcal{A}_{2}\mathcal{A}_{3}}U^\dagger_1
    +U_2\sigma_{\mathcal{A}_{1}\mathcal{A}_{2}\mathcal{B}_{2}\mathcal{A}_{3}}U^\dagger_2
    +U_3\sigma_{\mathcal{A}_{1}\mathcal{A}_{2}\mathcal{A}_{3}\mathcal{B}_{3}}U^\dagger_3)
\end{equation}
where $U_k$ are isometries of the form $U_1\ket{ij}_{\mathcal{A}_{2}\mathcal{A}_{3}}=\ket{iijj}_{\mathcal{A}_{2}\mathcal{B}_{2}\mathcal{A}_{3}\mathcal{B}_{3}}$, $U_2\ket{ij}_{\mathcal{A}_{1}\mathcal{A}_{3}}=\ket{iijj}_{\mathcal{A}_{1}\mathcal{B}_{1}\mathcal{A}_{3}\mathcal{B}_{3}}$ and $U_3\ket{ij}_{\mathcal{A}_{1}\mathcal{A}_{2}}=\ket{iijj}_{\mathcal{A}_{1}\mathcal{B}_{1}\mathcal{A}_{2}\mathcal{B}_{2}}$.

With this we can finally rewrite the two copies of the original state as
\begin{equation}
    \rho(p)^{\otimes2}=\,(1-2p)^2\rho_{\mathrm{diag}}+p(3-7p)\Gamma_1 
    +p(1-p)\Gamma_2+4p^2\Sigma,
\end{equation}
where $\rho_{\mathrm{diag}}$ is a normalized diagonal matrix.
With $m$ defined as~\eqref{eq:m},
the matrix $64(1-2p)^2\rho_{\mathrm{diag}}$ has the following entries: 

\begin{align*}
    &\underline{m} &&\hspace*{-3mm}  \underline{\rho_{\mathrm{diag}}(m,m)}\\
    & 1,22,43,64: &&\hspace*{-3mm}  (1-p)^2,\\
    & 2, 3, 5, 6, 9, 11, 17, 18, 21, 24, 30, 32, 33,\\
    & 35, 41, 44, 47, 48, 54, 56, 59, 60, 62, 63:&&  \hspace*{-3mm}1-10/3p+ 7/3p^2,\\
    & 4, 13, 16, 23, 26, 27, 38, 39, 42, 49, 52, 61: &&\hspace*{-3mm} 1-2p-13/3p^2,\\
    & 7, 8, 10, 12, 14, 15, 19, 20, 25, 28, 29, 31,\\
    & 34,36, 37, 40, 45, 46, 50, 51, 53, 55, 57, 58: &&\hspace*{-2mm} 1-6p+31/3p^2. 
\end{align*}

The terms $(1-p)^2$ and $1-6p+31/3p^2$ are positive for all values of $p$. The term $1-10/3p+ 7/3p^2$ is non-negative for $p\le3/7$ and $p\ge1$ and finally the term $1-2p-13/3p^2$ is non-negative for $(3-4\sqrt{3})/13\le p\le(3+4\sqrt{3})/13$. With this we have found a biseparable decomposition for all values $-1/7\le p\le(3+4\sqrt{3})/13$. From the bound shown in the main text
\begin{align}
    p   &>\,p\suptiny{0}{0}{(2)}_{\mathrm{GME}}(N)\,\coloneqq\,\frac{\sqrt{2^{N-1}-1}}{2^{N-1}+\sqrt{2^{N-1}-1}},
\end{align}
we know that all values above this bound are already GME\@.

\section{PPT criterion for isotropic GHZ states}\label{appendix:PPT criterion for isotropic GHZ states}

The isotropic GHZ states defined in the main text can be rewritten as
\begin{align}
    \rho(p) &=p\ket{\Phi^{+}}\!\!\bra{\Phi^{+}}_{\mathcal{A}_{1}\tilde{\mathcal{A}}_{2}}
    \!+\!\tfrac{1-p}{2^{N}}\,\mathds{1}_{\!\mathcal{A}_{1}}\hspace*{-3pt}\otimes\!
    \mathds{1}_{\!\tilde{\mathcal{A}}_{2}}
    \!+\!\tfrac{1-p}{2^{N}}\,\mathds{1}_{\!\mathcal{A}_{1}}\hspace*{-3pt}\otimes\!\mathds{1}_{\!\tilde{\mathcal{A}}_{2}^{\perp}}\nonumber\\[1mm]
    &=\,
    \tfrac{1+p}{2}\,\tilde{\rho}_{\mathcal{A}_{1}\tilde{\mathcal{A}}_{2}}\,
    +\tfrac{1-p}{2}\,\tfrac{1}{2^{N-1}}\,
    \mathds{1}_{\mathcal{A}_{1}}\hspace*{-3pt}\otimes\!
    \mathds{1}_{\tilde{\mathcal{A}}_{2}^{\perp}},
    \label{eq:GHZ with white noise rewritten appendix}
\end{align}
where $\ket{\Phi^{+}}_{\mathcal{A}_{1}\tilde{\mathcal{A}}_{2}}=\tfrac{1}{\sqrt{2}}\bigl(\ket{0}_{\mathcal{A}_{1}}\ket{\tilde{0}}_{\tilde{\mathcal{A}}_{2}}+\ket{1}_{\mathcal{A}_{1}}\ket{\tilde{1}}_{\tilde{\mathcal{A}}_{2}}\bigr)$ with $\ket{\tilde{i}}_{\tilde{\mathcal{A}}_{2}}=\bigotimes_{j=2}^{N}\ket{i}_{\mathcal{A}_{j}}$ for $i=0,1$, $\mathds{1}_{\!\tilde{\mathcal{A}}_{2}}=\sum_{i=0,1}\ket{\tilde{i}}\!\!\bra{\tilde{i}}$ and $\mathds{1}_{\!\tilde{\mathcal{A}}_{2}^{\perp}}=\mathds{1}_{2^{N-1}}-\mathds{1}_{\!\tilde{\mathcal{A}}_{2}}$.
We are now interested in checking for which values of $p$ the partial transpose of the two-qubit state $\tilde{\rho}_{\mathcal{A}_{1}\tilde{\mathcal{A}}_{2}}$ is positive semi-definite. Since the normalisation is irrelevant for this calculation, we can instead consider the partial transpose of the unnormalised operator $\tfrac{1+p}{2}\,\tilde{\rho}_{\mathcal{A}_{1}\tilde{\mathcal{A}}_{2}}$ whose partial transpose is given by
\begin{align}
    \bigl(\tfrac{1+p}{2}\,\tilde{\rho}_{\mathcal{A}_{1}\tilde{\mathcal{A}}_{2}}\bigr)^{T_{\tilde{\mathcal{A}}_{2}}} &=\,
    \begin{pmatrix}
        \tfrac{p}{2}+\tfrac{1-p}{2^{N}} & 0 & 0 & 0 \\
        0 & \tfrac{1-p}{2^{N}}  & \tfrac{p}{2} & 0 \\
        0 & \tfrac{p}{2} & \tfrac{1-p}{2^{N}} & 0 \\
        0 & 0 & 0 & \tfrac{p}{2}+\tfrac{1-p}{2^{N}}
    \end{pmatrix}.
    \label{eq:two by two block}
\end{align}
The only potentially negative eigenvalue of this matrix is $(1-p)/2^{N}-p/2$ and we hence find that $\tilde{\rho}_{\mathcal{A}_{1}\tilde{\mathcal{A}}_{2}}$ is positive semi-definite for $p\leq p_{\mathrm{crit}}\coloneqq 1/(1+2^{N-1})$. Since $\tilde{\rho}_{\mathcal{A}_{1}\tilde{\mathcal{A}}_{2}}$ is a two-qubit state, the PPT criterion is necessary and sufficient for separability, and the state $\rho(p)$ hence has a separable decomposition with respect to the bipartition $\mathcal{A}_{1}|\mathcal{A}_{2}\ldots \mathcal{A}_{N}$ for $p\leq p_{\mathrm{crit}}$.

Since $\rho(p)$ is invariant under exchanges of any qubits, this separability threshold applies for any bipartition of separating any one qubit from the remaining $N-1$ qubits. Moreover, it is easy to see that the arguments presented above hold also for any bipartition into $M$ and $N-M$ qubits by choosing suitable single-qubit subspaces in both the $M$-qubit and $(M-N)$-qubit Hilbert spaces.

We also note that the threshold value $p_{\mathrm{crit}}$ for partition-separability trivially coincides with the PPT threshold for any chosen bipartition of $\rho(p)$ because the only non-diagonal $2\times2$-block of the partial transpose is always of the form of the right-hand side of Eq.~(\ref{eq:two by two block}). In particular, this implies that all states $\rho(p)$ are non-PPT (NPT) entangled across any bipartition for $p> p_{\mathrm{crit}}$ and separable below this value. Consequently, there are no PPT entangled isotropic GHZ states.

\section{PPT-triangle states and GME activation}\label{appendix:PPT entangled GME activation}

To investigate whether multi-copy GME activation is possible from bound entanglement, we first consider a biseparable three-party state with no distillable bipartite entanglement across any bipartition; i.e., the state is positive under partial transposition across all cuts. Since the set of PPT states is convex, we may construct such a state as a convex combination of terms where one party is uncorrelated with the others, while the remaining two parties share a PPT entangled state, i.e.,
\begin{equation}
\label{eq:biseparble_ppt}
    \rho_{\mathcal{A}_1\mathcal{A}_2\mathcal{A}_3}
    =\, p_1 \rho_{\mathcal{A}_1}\otimes  \rho\suptiny{0}{0}{\mathrm{PPT}}_{\mathcal{A}_2\mathcal{A}_3}
    +p_2\rho_{\mathcal{A}_2}\otimes  \rho\suptiny{0}{0}{\mathrm{PPT}}_{\mathcal{A}_1\mathcal{A}_3}
    +\,p_3\rho_{\mathcal{A}_3}\otimes 
     \rho\suptiny{0}{0}{\mathrm{PPT}}_{\mathcal{A}_1\mathcal{A}_2},
\end{equation}
where $\sum_{i}p_i=1$, $p_i\geq 0$ and $ \rho\suptiny{0}{0}{\mathrm{PPT}}_{\mathcal{A}_i\mathcal{A}_j}$ for $i,j\in\{1,2,3\}$ are PPT entangled states. Here we note that the existence of such a decomposition guarantees biseparability, but it does not a priori rule out that such a state may be partition-separable (or even fully separable). If $\rho_{\mathcal{A}_1\mathcal{A}_2\mathcal{A}_3}$ is separable with respect to one or several of the bipartitions $\mathcal{A}_1|\mathcal{A}_2\mathcal{A}_3$, $\mathcal{A}_1\mathcal{A}_2|\mathcal{A}_3$ and $\mathcal{A}_2|\mathcal{A}_1\mathcal{A}_3$, then GME activation is not possible for any number of copies. However, as we show here, for certain choices of the $ \rho\suptiny{0}{0}{\mathrm{PPT}}_{\mathcal{A}_i\mathcal{A}_j}$ and $\rho_{\mathcal{A}_k}$, three copies of $\rho_{\mathcal{A}_1\mathcal{A}_2\mathcal{A}_3}$ are GME, which thus also shows that the single-copy states in question are not partition-separable (or fully separable).

To continue, let us consider the particular situation where each of the three parties $\mathcal{A}_{i}$ for $i=1,2,3$ consists of three subsystems $\mathcal{A}_{i}\suptiny{0}{-2}{(j)}$ for $j=1,2,3$. In this situation, a particular example for a state of the form of Eq.~(\ref{eq:biseparble_ppt}) is given by
\begin{align}
     \rho_{\mathcal{A}_{1}\mathcal{A}_{2}\mathcal{A}_{3}}
     &=\hspace*{-7pt}
     \sum\limits_{\substack{i=1,2,3\\ i\neq j\neq k\neq i \\ j<k}}
     \hspace*{-5pt}
     p_{i}\ \rho_{\mathcal{A}_{i}\suptiny{0}{-2}{(i)}}\otimes
      \rho\suptiny{0}{0}{\mathrm{PPT}}_{\mathcal{A}_{j}\suptiny{0}{-2}{(i)}\mathcal{A}_{k}\suptiny{0}{-2}{(i)}}
     \otimes\!\bigotimes\limits_{\substack{m,n=1\\ n\neq i}}^{3}\ket{0}\!\!\bra{0}_{\mathcal{A}_{m}\suptiny{0}{-2}{(n)}},\nonumber
\end{align}
where the states $ \rho\suptiny{0}{0}{\mathrm{PPT}}_{\mathcal{A}_{j}\suptiny{0}{-2}{(i)}\mathcal{A}_{k}\suptiny{0}{-2}{(i)}}$ are PPT entangled states that will be specified later. Now, suppose that three copies, $\rho_{\mathcal{A}_1\mathcal{A}_2\mathcal{A}_3}$, $\rho_{\mathcal{B}_1\mathcal{B}_2\mathcal{B}_3}$, and $\rho_{\mathcal{C}_1\mathcal{C}_2\mathcal{C}_3}$, are shared. By projecting the subsystems $\mathcal{A}_{1}\suptiny{0}{-2}{(1)}$ of the first copy, $\mathcal{B}_{2}\suptiny{0}{-2}{(2)}$ of the second copy, and $\mathcal{C}_{3}\suptiny{0}{-2}{(3)}$ of the third copy into the subspaces orthogonal to the states $\ket{0}_{\mathcal{A}_{1}\suptiny{0}{-2}{(1)}}$, $\ket{0}_{\mathcal{B}_{2}\suptiny{0}{-2}{(2)}}$, and $\ket{0}_{\mathcal{C}_{3}\suptiny{0}{-2}{(3)}}$, respectively,
the three parties can (deterministically) prepare the states $ \rho\suptiny{0}{0}{\mathrm{PPT}}_{\mathcal{A}_{2}\suptiny{0}{-2}{(1)}\mathcal{A}_{3}\suptiny{0}{-2}{(1)}}$, $ \rho\suptiny{0}{0}{\mathrm{PPT}}_{\mathcal{B}_{1}\suptiny{0}{-2}{(2)}\mathcal{B}_{3}\suptiny{0}{-2}{(2)}}$, and $ \rho\suptiny{0}{0}{\mathrm{PPT}}_{\mathcal{C}_{1}\suptiny{0}{-2}{(3)}\mathcal{C}_{2}\suptiny{0}{-2}{(3)}}$.
All other subsystems can be discarded.
Consequently, three copies of $\rho_{\mathcal{A}_{1}\mathcal{A}_{2}\mathcal{A}_{3}}$ allow the parties to establish a state of the form
\begin{align}
\label{eq:ppt_triangle}
    \rho^{\mathrm{\Delta PPT}}_{\mathcal{O}_{1}\mathcal{O}_{2}\mathcal{O}_{3}}
    &
    \coloneqq \rho\suptiny{0}{0}{\mathrm{PPT}}_{\mathcal{A}_2\mathcal{A}_3}\otimes \rho\suptiny{0}{0}{\mathrm{PPT}}_{\mathcal{B}_1\mathcal{B}_3}\otimes \rho\suptiny{0}{0}{\mathrm{PPT}}_{\mathcal{C}_1\mathcal{C}_2}
\end{align}
via LOCC\@. For ease of notation we have dropped the superscripts identifying the particular subsystems, e.g., using the label $\mathcal{A}_{i}$ instead of $\mathcal{A}_{i}\suptiny{0}{-2}{(j)}$.
We call a state in this form a \textit{PPT-triangle} state, where the parties $1$, $2$, and $3$ have access to systems $\mathcal{B}_1 \mathcal{C}_1$, $\mathcal{A}_2 \mathcal{C}_2$, and $\mathcal{A}_3 \mathcal{B}_3$, respectively.
We further note that every such PPT-triangle state can be created via LOCC from three copies of a biseparable state of the form of Eq.~(\ref{eq:biseparble_ppt}).

Therefore, we reach the following claim:
if there is a GME state that is PPT-triangle, then multi-copy GME activation is achievable for (some) biseparable states that are PPT across every cut.
Consequently, the problem reduces to proving the existence of a PPT-triangle state that exhibits GME\@.
To find such a state, we construct a one-parameter family of two-qutrit states given by
\begin{align}
\label{eq:ppt_triangle_example1}
     \rho\suptiny{0}{0}{\mathrm{PPT}}_{\mathcal{X}\mathcal{Y}}(p)
    \coloneqq\tfrac{1}{\mathcal{N}_p}\bigl[&(|00\rangle\nl+\nl|11\rangle\nl+\nl|22\rangle)(\langle00|\nl+\nl\langle11|\nl+\nl\langle22|) 
    +p(|01\rangle\langle01|+|12\rangle\langle12|+|20\rangle\langle20|)\nonumber\\
    +\tfrac{1}{p}&(|02\rangle\langle02|+|10\rangle\langle10|+|21\rangle\langle21|)\bigr],
\end{align}
for all $p>0$, where $\mathcal{X}$ and $\mathcal{Y}$ labels the first and second qutrit, respectively, $\mathcal{N}_p=3(1+p+\tfrac{1}{p})>0$ is a normalization constant. The partial transpose of $ \rho\suptiny{0}{0}{\mathrm{PPT}}_{\mathcal{X}\mathcal{Y}}(p)$ has eigenvalues $\lambda_{1}=0$, $\lambda_{2}=\mathcal{N}_p>0$, and $\lambda_{3}=\mathcal{N}_p(p+\tfrac{1}{p})>0$, each thrice degenerate, and $ \rho\suptiny{0}{0}{\mathrm{PPT}}_{\mathcal{X}\mathcal{Y}}(p)$ is hence PPT\@.
We can then choose the PPT states in Eq.~(\ref{eq:ppt_triangle}) from this family of two-qutrit states, such that
\begin{align}
\label{eq:ppt_triangle_example2}
    \rho^{\mathrm{\Delta PPT}}_{\mathcal{O}_{1}\mathcal{O}_{2}\mathcal{O}_{3}}(x,y,z)\coloneqq
     \rho\suptiny{0}{0}{\mathrm{PPT}}_{\mathcal{A}_2\mathcal{A}_3}(x)
    \otimes \rho\suptiny{0}{0}{\mathrm{PPT}}_{\mathcal{B}_1\mathcal{B}_3}(y)
    \otimes \rho\suptiny{0}{0}{\mathrm{PPT}}_{\mathcal{C}_1\mathcal{C}_2}(z).
\end{align}
To show that the state is GME with respect to the partition $\mathcal{O}_{1}|\mathcal{O}_{2}|\mathcal{O}_{3}$ it suffices to detect GME between subspaces $\mathcal{D}_{1}$, $\mathcal{D}_{2}$, and $\mathcal{D}_{3}$ of $\mathcal{O}_{1}$, $\mathcal{O}_{2}$, and $\mathcal{O}_{3}$, respectively. Specifically, we consider the single-qutrit subspaces $\mathcal{D}_{1}$, $\mathcal{D}_{2}$, and $\mathcal{D}_{3}$ spanned by $\{|ii\rangle_{\mathcal{B}_1\mathcal{C}_1}\}_{i=0,1,2}$, $\{|jj\rangle_{\mathcal{A}_2\mathcal{C}_2}\}_{j=0,1,2}$, and $\{|kk\rangle_{\mathcal{A}_3\mathcal{B}_3}\}_{k=0,1,2}$, respectively, and thus the projection of $\rho^{\mathrm{\Delta PPT}}_{\mathcal{O}_{1}\mathcal{O}_{2}\mathcal{O}_{3}}(x,y,z)$ onto the three-qutrit subspace spanned by $\{|ii\rangle_{\mathcal{B}_1\mathcal{C}_1}\otimes|jj\rangle_{\mathcal{A}_2\mathcal{C}_2}\otimes|kk\rangle_{\mathcal{A}_3\mathcal{B}_3}\}_{i,j,k=0,1,2}$.  We let $\rho^{\mathrm{\Delta PPT}}_{\mathcal{D}_{1}\mathcal{D}_{2}\mathcal{D}_{3}}(x,y,z)$ denote the resulting state.

To this state, we apply a three-party GME witness $W_3$ (see~\cite[example 2]{HuberSengupta2014}) based on the lifted Choi-map witnesses from~\cite{ClivazHuberLamiMurta2017} of the form
\begin{align}
    W_3=\ \ &\ketbra{000}{000}+\ketbra{001}{001}+\ketbra{011}{011}+\ketbra{020}{020}+\ketbra{101}{101}+\ketbra{111}{111}+\ketbra{112}{112}\\
    +\,&\ketbra{122}{122}+\ketbra{200}{200}+\ketbra{212}{212}+\ketbra{220}{220}+\ketbra{222}{222}-\ketbra{000}{111}-\ketbra{000}{222}\notag\\
    -\,&\ketbra{111}{222}-\ketbra{111}{000}-\ketbra{222}{000}-\ketbra{222}{111}.\notag
\end{align}

Applying it to our state yields the expression
\begin{align}
    \tr[W_3\,\rho^{\mathrm{\Delta PPT}}_{\mathcal{D}_{1}\mathcal{D}_{2}\mathcal{D}_{3}}(x,y,z)]
    &=\tfrac{3}{\mathcal{N}_x\mathcal{N}_y\mathcal{N}_z}(xy\nl+\nl\tfrac{z}{x}\nl+\nl yz-1).
\end{align}
We see that for certain values of $x$, $y$ and $z$ the expected value of the witness can be negative, e.g., for states of the form $\rho^{\mathrm{\Delta PPT}}_{\mathcal{D}_{1}\mathcal{D}_{2}\mathcal{D}_{3}}(1,y,y)$ with $0<y<\sqrt{2}-1$, thus detecting GME in this range.

Finally, an observation that we can make about the PPT-triangle states in Eq.~(\ref{eq:ppt_triangle_example2}) is that the third tensor factor $ \rho\suptiny{0}{0}{\mathrm{PPT}}_{\mathcal{C}_1\mathcal{C}_2}(z)$ is not even necessary to obtain GME\@. Indeed, the state
\begin{align}
\label{eq:ppt_triangle_example3}
    \rho^{\mathrm{\wedge PPT}}_{\mathcal{A}_2\mathcal{A}_3\mathcal{B}_1\mathcal{B}_3}(x,y)= \rho\suptiny{0}{0}{\mathrm{PPT}}_{\mathcal{A}_2\mathcal{A}_3}(x)\otimes \rho\suptiny{0}{0}{\mathrm{PPT}}_{\mathcal{B}_1\mathcal{B}_3}(y)
\end{align}
is GME for certain values of $x$ and $y$.
To show this it again suffices detecting GME on a subspace. Consider the projection onto the three-qutrit subspace spanned by $\{\otimes|i\rangle_{\mathcal{B}_1}\otimes|j\rangle_{\mathcal{A}_2}|kk\rangle_{\mathcal{A}_3\mathcal{B}_3}\}_{i,j,k=0,1,2}$ and denote the resulting state by $\rho^{\mathrm{\wedge PPT}}_{\mathcal{D}_1\mathcal{D}_2\mathcal{D}_3}(x,y)$. With the same witness $W_3$ as before we obtain
\begin{align}
    \tr[W_3\, \rho^{\mathrm{\wedge PPT}}_{\mathcal{D}_1\mathcal{D}_2\mathcal{D}_3}(x,y)]=\tfrac{3}{\mathcal{N}_x\mathcal{N}_y}\, (x+y+xy-1).
\end{align}
For instance, for $x=y<\sqrt{2}-1$, this expression becomes negative, thus detecting GME\@. We can thus conclude that PPT entanglement across two out of the three cuts and two copies of the original state are already enough for GME activation.

\end{document}